\documentclass[prd, superscriptaddress,nofootinbib]{revtex4}

\begin{document}
%


\title{Reply to the comment by Szpak on leading order asymptotics \\of late-time tails for a
self-gravitating massless scalar field}
\author{Piotr Bizo\'n}
\affiliation{M. Smoluchowski Institute of Physics, Jagiellonian
University, Krak\'ow, Poland}
\author{Tadeusz Chmaj}
\affiliation{H. Niewodniczanski Institute of Nuclear
   Physics, Polish Academy of Sciences,  Krak\'ow, Poland}
   \affiliation{Cracow University of Technology, Krak\'ow,
    Poland}
\author{Andrzej Rostworowski}
\affiliation{M. Smoluchowski Institute of Physics, Jagiellonian
University, Krak\'ow, Poland}

\date{\today}
\begin{abstract}
We respond to N. Szpak's comment [arXiv: 0907.5146v2] on our paper "Late-time tails of a
self-gravitating massless scalar field, revisited", Class. Quantum Grav. \textbf{26}, 175006
(2009).
\end{abstract}

\maketitle
In a recent paper \cite{bcr5} we revisited the problem of long-time behavior of a spherically
symmetric self-gravitating massless scalar field.  We emphasized that the asymptotic convergence
to a static equilibrium (Minkowski or Schwarzschild) is an essentially nonlinear phenomenon which
cannot, despite many assertions to the contrary in the literature, be properly described by the
theory of linearized perturbations on a fixed static asymptotically flat background (Price's
tails). To substantiate this claim in the case of small initial data we computed the late-time
tails (both the decay rate and the amplitude) in four and higher even spacetime dimensions using
nonlinear perturbation theory and we verified the results numerically. The reason for considering
this problem in higher dimensions was motivated by the desire to demonstrate an accidental and
misleading character of equality of decay rates of linear and nonlinear tails in four dimensions.

Recently, Szpak \cite{szpak} showed that most (but not all) of our results can be reproduced by
an asymptotic  calculation which keeps track of only leading order terms in the perturbation
equations. This is an interesting observation which provides insight into the mechanism of some
cancelations in the asymptotic expansions. Delighted as we are to see someone going through the
details of our calculations (and confirming the results), we wish to make a couple of clarifying
remarks:
\begin{itemize}
\item From a practical viewpoint, the calculation by Szpak is only superficially more efficient.
Although his asymptotic equations are simpler to deal with, this fact is counterbalanced by an
extra effort which goes into proving that the terms dropped from the equations are really
irrelevant.

\item Using only leading order asymptotics
 Szpak could not  obtain our
interpolating formula (Eq.(41) in \cite{bcr5}) for the tail in $d+1$ dimensions (for odd $d\geq
5$). Note that this formula contains significantly more information about the asymptotic behavior
of solutions than its limiting cases at time and null infinities. (Nota bene, the first version
of Szpak's comment [arXiv: 0907.5146v1] contained a critical remark about the validity of our
results at null infinity; this remark, based on a misunderstanding, was withdrawn from the final
version of the comment so we are glad not to have to  respond to it.)

\item The identification of a dominant term in the perturbation equations is useful only if this
term gives rise to a nonzero tail in the leading order (as it luckily happens in the case at
hand).
 If there were
further cancelations, it would be impossible to see them at the level of  equations and an
asymptotic calculation of \cite{szpak} would not work. Such a situation arises, for example, for
self-gravitating wave maps \cite{bcr6}.

\end{itemize}

\end{document}